\documentclass[pra,twocolumn,letterpaper,superscriptaddress]{revtex4}
\usepackage{graphicx,amsmath,amssymb,amsfonts,latexsym,color,dcolumn,bm,epsfig}
\usepackage[latin1]{inputenc}

\def\bbm[#1]{\mbox{\boldmath $#1$}}

\newcommand{\TE}{\text{TE}}
\newcommand{\TM}{\text{TM}}

\begin{document}

\title{Graphene-based photovoltaic cells for near-field thermal energy conversion}

\author{Riccardo Messina}\email{riccardo.messina@institutoptique.fr}
\author{Philippe Ben-Abdallah}\email{pba@institutoptique.fr}
\affiliation{Laboratoire Charles Fabry, UMR 8501, Institut d'Optique, CNRS, Universit\'{e} Paris-Sud 11, 2, Avenue Augustin Fresnel, 91127 Palaiseau Cedex, France.}

\date{\today}

\begin{abstract}
Thermophotovoltaic devices are energy-conversion systems generating an electric current from the thermal photons radiated by a hot body. In far field, the efficiency of these systems is limited by the thermodynamic Schockley-Queisser limit corresponding to the case where the source is a black body. On the other hand, in near field, the heat flux which can be transferred to a photovoltaic cell can be several orders of magnitude larger because of the contribution of evanescent photons. This is particularly true when the source supports surface polaritons. Unfortunately, in the infrared where these systems operate, the mismatch between the surface-mode frequency and the semiconductor gap reduces drastically the potential of this technology. Here we show that graphene-based hybrid photovoltaic cells can significantly enhance the generated power paving the way to a promising technology for an intensive production of electricity from waste heat.
\end{abstract}

\maketitle

A hot body at temperature $T$ radiates an electromagnetic field in its surroundings because of local thermal fluctuations. In the close vicinity of its surface, at distances smaller than the thermal wavelength $\lambda_{th}=\hbar c/(k_{B}T)$, the electromagnetic energy density is several orders of magnitude larger than in far field \cite{JoulainSurfSciRep05,VolokitinRevModPhys07}. Hence, the near-field thermal radiation associated to non-propagating photons which remain confined on the surface is a potentially important source of energy. By approaching a PV cell \cite{BasuIntJEnergyRes07} in proximity of a thermal emitter, this energy can be extracted by photon tunneling toward the cell. Such devices, also called near-field thermophotovoltaic (NTPV) systems, have been proposed ten years ago \cite{DiMatteoApplPhysLett01}. In presence of resonant surface modes such as surface polaritons, the flux exchanged in near-field between source and photodiode drastically overcomes the propagative contribution \cite{KittelPRL05,HuApplPhysLett08,ShenNanoLetters09,KralikRevSciInstrum11,OttensPRL11}. This discovery has opened new possibilities for the development of innovative technologies for nanoscale thermal management, heating-assisted data storage \cite{SrituravanichNanoLett04}, IR sensing and spectroscopy \cite{DeWildeNature06,JonesNanoLetters12} and has paved the way to a new generation of NTPV energy-conversion devices \cite{NarayanaswamyApplPhysLett03,Laroche1}.

Despite its evident interest, several problems still limit the technological development of NTPV conversion. The main one is the mismatch between the frequency of surface polaritons supported by the hot source and the gap frequency of the cell (typically a semiconductor). Indeed, all photons with energy larger than the frequency gap are not totally converted into hole-electron pair but a part of their energy is dissipated via phonon excitation. Besides, low-energy photons do not contribute to the production of electricity but are only dissipated into heat within the atomic lattice. To overcome this problem, we introduce here a \emph{relay} between the source and the cell to make the transport of heat more efficient. Graphene is a natural candidate to carry out this function. Indeed, this two-dimensional monolayer of carbon atoms which has proved to be an extremely surprising material with unusual electrical and optical properties \cite{Geim1,Geim2,Avouris} can be tailored by modifying the chemical potential 
to be resonant between the gap frequency of the semiconductor and the resonance frequency of the polariton supported by the source. In the context of heat transfer, the role of graphene has been recently investigated \cite{PerssonJPhysCondensMatter10,VolokitinPRB11,SvetovoyPRB12,IlicPRB12}, confirming its tunability and paving the way to promising thermal devices such as thermal transistors. Furthermore, a NTPV cell in which a suspended graphene sheet acts as source has been recently considered \cite{IlicOptExpress12}. We propose here a modification of the standard NTPV scheme, in which the surface of the cell is covered with a graphene sheet. As we will show, this enables to exploit at the same time the existence of a surface phonon-polariton of the source and the tunability of graphene as an efficient tool to enhance the source-cell coupling.

\begin{figure}[h!]
\scalebox{0.3}{\includegraphics{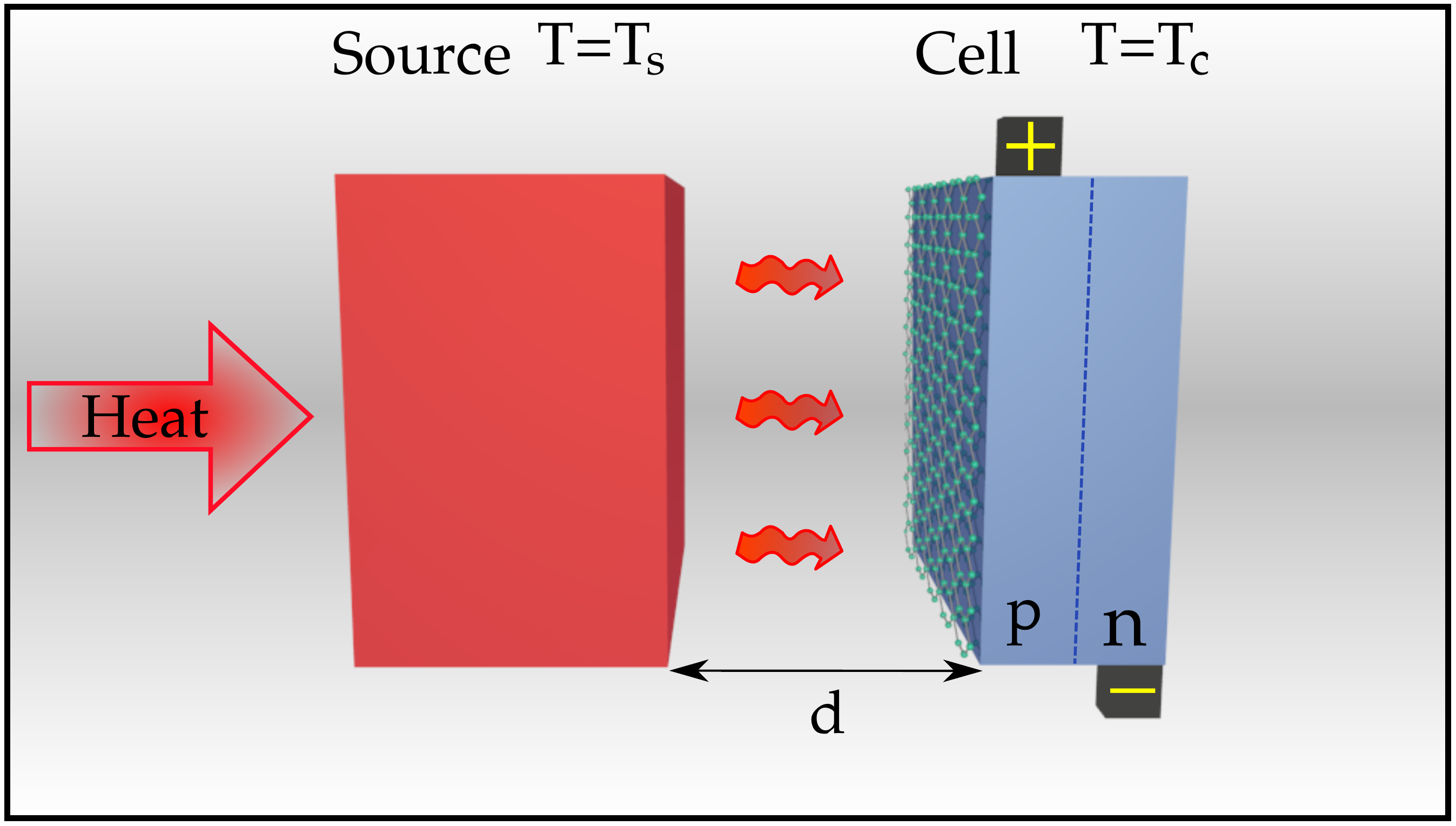}}
\caption{Scheme of a typical NTPV cell. A hot source (temperature $T_s$) is placed in front of a cell at temperature $T_c$, which is typically a p-n junction. The source is heated by an external radiation flow, and the temperature of the cell is kept constant in time. The radiation flux between source and cell is converted into an electrical current inside the cell, extracted by means of the two electrodes connected to the junction.}
\label{FigGeometry}\end{figure}

Figure \ref{FigGeometry} outlines our novel hybrid graphene-semiconductor NTPV system. A hot source made of hexagonal Boron Nitride (hBN) at temperature $T_s=450\,$K, eventually heated by an external primary source, is placed in the proximity of a graphene-covered cell made of Indium Antimonide (InSb) at temperature $T_c=300\,$K, having gap frequency $\omega_g\simeq2.583\times10^{14}\,\text{rad s}^{-1}$. In the frequency domain of interest, the InSb dielectric permittivity can be described using the simple model \cite{IlicOptExpress12}
\begin{equation}\label{EpsCell}\begin{split}\varepsilon_2(\omega)&=\Bigl(n_r(\omega)+i\frac{c\alpha(\omega)}{2\omega}\Bigr)^2\\
\alpha(\omega)&=\begin{cases}0 & \omega<\omega_g\\\alpha_0\sqrt{\frac{\omega-\omega_g}{\omega_g}} & \omega>\omega_g\end{cases}\end{split}\end{equation}
where $n_r(\omega)$ is the refractive index and the choice $\alpha_0=0.7\,\mu\text{m}^{-1}$ reasonably reproduces the experimental values of absorption \cite{GobeliPhysRev60}. As discussed before, it is important to choose a source having a strong emission, and possibly a surface phonon-polariton resonance, at frequencies slightly larger than the gap frequency $\omega_g$. For our purpose we have chosen hBN which we describe (ignoring for the sake of simplicity its anisotropy) using a Drude-Lorentz model $\epsilon_1(\omega)=\varepsilon_\infty(\omega^2-\omega_L^2+i\Gamma\omega)/(\omega^2-\omega_R^2+i\Gamma\omega)$ with $\varepsilon_\infty=4.88$, $\omega_L=3.032\times10^{14}\,\text{rad s}^{-1}$, $\omega_R=2.575\times10^{14}\,\text{rad s}^{-1}$ and $\Gamma=1.001\times10^{12}\,\text{rad s}^{-1}$\cite{NarayanaswamyApplPhysLett03}. This model predicts the existence of a surface phonon-polariton resonance at frequency $\omega_\text{spp}\simeq2.960\times10^{14}\,\text{rad s}^{-1}$, larger than $\omega_g$ as wished.

The hybrid configuration will be compared to the case without graphene sheet. The optical properties of graphene are accounted for by means of a 2D frequency-dependent conductivity (see Appendix for details) \cite{FalkovskyJPhysConfSer08} as already done in the context of a heat-transfer calculations in \cite{IlicOptExpress12,IlicPRB12}. We are going to study the modifications to the radiation flux between the source and the cell, as well as to the power produced by the device, due to the presence of graphene. In particular, our main scope is to investigate whether the presence of graphene is able to enhance the efficiency of the NTPV cell and possibly the output power as well.

\begin{figure}[h!]
\scalebox{0.05}{\includegraphics{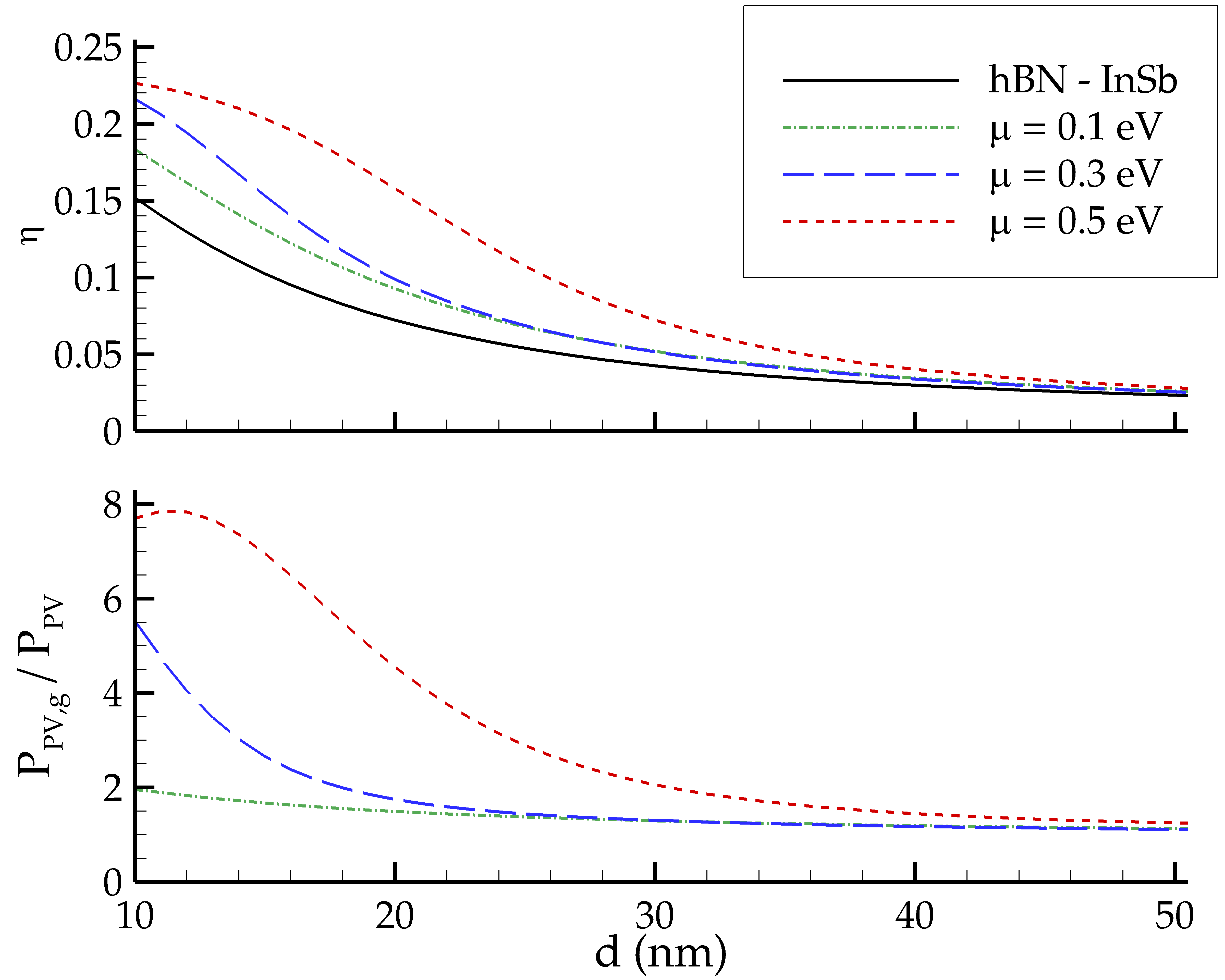}}
\caption{Efficiency and ratio of produced electric power in presence of graphene. The curves correspond to three different values of chemical potential of graphene. For any distance $d$ here represented, the presence of the graphene sheet produces an enhancement of efficiency as well as an amplification of electric power. The overall effect decreases with respect to the source-cell distance $d$.}
\label{FigEtaPPV}\end{figure}

First, we need a general expression for the heat flux exchanged between two planar surfaces. This distance-dependent flux $\varphi(d)$ can be expressed in near-field regime as $\varphi(d)=\int_0^{+\infty}\frac{d\omega}{2\pi}\phi(\omega,d)$, where
\begin{equation}\label{Flux}\phi(\omega,d)=\hbar\omega\,n_{sc}(\omega)\sum_p\int_{ck>\omega}\frac{d^2\mathbf{k}}{(2\pi)^2}\mathcal{T}_p(\omega,\mathbf{k},d).\end{equation}
This expression implies the sum over all the evanescent modes of the electromagnetic field (identified by the frequency $\omega$, the transverse wavevector $\mathbf{k}=(k_x,k_y)$ and the polarization $p$ taking the values $p=\text{TE}$ and $p=\text{TM}$) of the product of the energy $\hbar\omega$ carried by each mode $(\omega,\mathbf{k},p)$, the difference $n_{sc}(\omega)=n(\omega,T_s)-n(\omega,T_c)$, $n(\omega,T)=(e^{\hbar\omega/k_BT}-1)^{-1}$ being the distribution function inside the reservoir of modes at temperature $T$, and a transmission probability $\mathcal{T}_p(\omega,\mathbf{k},d)$ through the separation gap assuming values between 0 and 1. In this Landauer-like decomposition \cite{BenAbdallahPRB10,BiehsPRL10,LandauerPhilosMag70,ButtikerPRL86} the transmission probability $\mathcal{T}_p(\omega,\mathbf{k},d)$ represents an absolute measure of the contribution of a given mode to the energy exchange. In the case of two semi-infinite parallel planar media this quantity reads \cite{BenAbdallahPRB10}
\begin{equation}\label{T2s}\mathcal{T}_p(\omega,\mathbf{k},d)=\frac{4\,{\rm Im}(r_{1p}){\rm Im}(r_{2p})e^{2ik_zd}}{|1-r_{1p}r_{2p}e^{2ik_zd}|^2},\end{equation}
where $k_z=\sqrt{\omega^2/c^2-\mathbf{k}^2}$ is the $z$ component of the wavevector in vacuum, while $r_{1p}$ and $r_{2p}$ are the ordinary vacuum-medium Fresnel coefficients corresponding to polarization $p$ and bodies 1 and 2 respectively. For a NTPV cell, the expression \eqref{Flux} has to be modified in order to take into account the fact that the cell is a direct-gap semiconductor. Hence, the radiative power exchanged between the source and the cell is given by
\begin{equation}\label{Prad}\begin{split}P_\text{rad}(d)&=\int_0^{+\infty}\frac{d\omega}{2\pi}\hbar\omega\,n(\omega,T_s)\phi(\omega)\\
&\,-\int_{\omega_g}^{+\infty}\frac{d\omega}{2\pi}\hbar\omega\,n(\omega-\omega_0,T_c)\phi(\omega,d),\end{split}\end{equation}
where $\omega_0=eV_0/\hbar$, $V_0$ being the potential difference at which the cell is operating, quantity for which we take a value slightly below the theoretical limit $\tilde{V}_0=\omega_g(1-T_c/T_s)$ \cite{IlicOptExpress12}. As for the expression of the electric power which is generated from this flux, it reads
\begin{equation}\label{PPV}\begin{split}P_\text{PV}(d)&=\int_{\omega_g}^{+\infty}\frac{d\omega}{2\pi}\hbar\omega_0\,n(\omega,T_s)\phi(\omega)\\
&\,-\int_{\omega_g}^{+\infty}\frac{d\omega}{2\pi}\hbar\omega_0\,n(\omega-\omega_0,T_c)\phi(\omega,d).\end{split}\end{equation}

As anticipated, the evolution of both $P_\text{PV}$ and the cell efficiency $\eta=P_\text{PV}/P_\text{rad}$ has to be followed when modifying a NTPV device, the challenge being their simultaneous enhancement. In Figure \ref{FigEtaPPV} we represent both the efficiency and the ratio of electric powers for four different configurations, namely without graphene and with a sheet deposed on the surface of the cell for three different values of the chemical potential $\mu$. Regarding the efficiency, we see that graphene produces indeed an enhancement. For example, at $d=16\,$nm $\eta$ goes from around 10\% in absence of graphene to almost 20\% for $\mu=0.5\,$eV, approaching considerably the ideal Carnot limit $1-T_c/T_s\simeq33\%$. We also see that for larger distances and for any choice of $\mu$ the graphene-modified efficiencies tend to the one associated to the standard hBN\,-\,InSb system. As for the electrical power, the presence of graphene produces an amplification going up to values of the order of 8, 
showing that both desired conditions are met by our modification scheme.

\begin{figure}[h!]
\scalebox{0.05}{\includegraphics{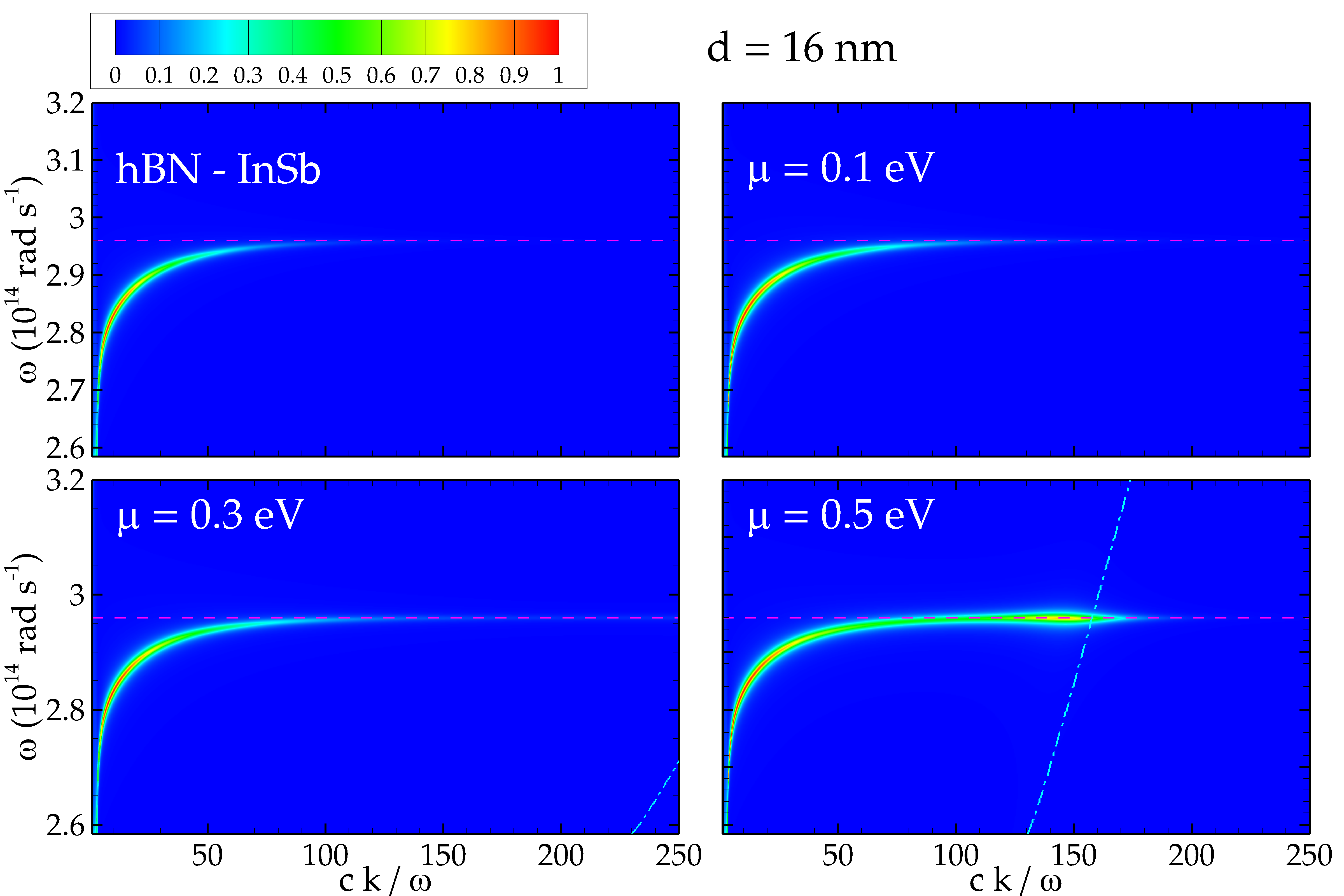}}
\caption{Transmission probability with and without graphene. The transmission probability \eqref{T2s} is represented in absence of graphene, and with graphene for $\mu=0.1,0.3,0.5\,$eV. The horizontal dashed line represents the frequency $\omega_\text{spp}$ of the surface phonon-polariton resonance of hBN, while the dot-dashed lines describe the resonance of the graphene\,-\,InSb system. The enhancement of transmission probability is clearly visible around the intersection point of the two branches for $\mu=0.5\,$eV, whereas the same coupling produces an increase of the cutoff wavevector $k_c$ for smaller values of the chemical potential.}
\label{FigFwk}\end{figure}

We now show in Figure \ref{FigFwk} the transmission probability $\mathcal{T}_p(\omega,\mathbf{k},d)$ for a source-cell distance of $d=16\,$nm, without and with graphene. We first observe that in absence of graphene the modes contributing to the effect are concentrated in proximity of the surface resonance $\omega_\text{spp}$ of hBN. This resonance branch decays with respect to the wavevector $k$, and in particular is no longer visible around $ck/\omega_\text{spp}\simeq100$. This cutoff, well-known in the theory of radiative heat transfer, is mainly connected to the distance between the bodies and is roughly given by $k_c\simeq1/d$ \cite{BenAbdallahPRB10,BiehsPRL10}. We remark that the InSb cell does not support any surface mode in the frequency region of interest. This is no longer true in presence of graphene. The modification of the optical properties due to the deposed sheet induces the appearance of a resonant surface mode associated to the graphene-modified cell, represented in figure by the dot-dashed 
line. This is not visible for $\mu=0.1\,$eV, appears for very high wavevectors for $\mu=0.3\,$eV, is clearly visible for the largest chosen value $\mu=0.5\,$eV. The new physical mechanism generated by the presence of graphene can be described in terms of a coupling between the two surface modes. For $\mu=0.5\,$eV this manifestly produces an enhanced region of transmission probability around the intersection point of the two branches, while for $\mu=0.1$ and 0.3\,eV the same coupling, taking place at larger values of $ck/\omega$, gives rise to an increase of the cutoff wavevector $k_c$. The fact that $k_c$ is a decreasing function of $d$ and that the graphene-induced resonant modes exist for high values of the wavevector explains why the efficiency enhancement decreases with $d$.

\begin{figure}[h!]
\scalebox{0.05}{\includegraphics{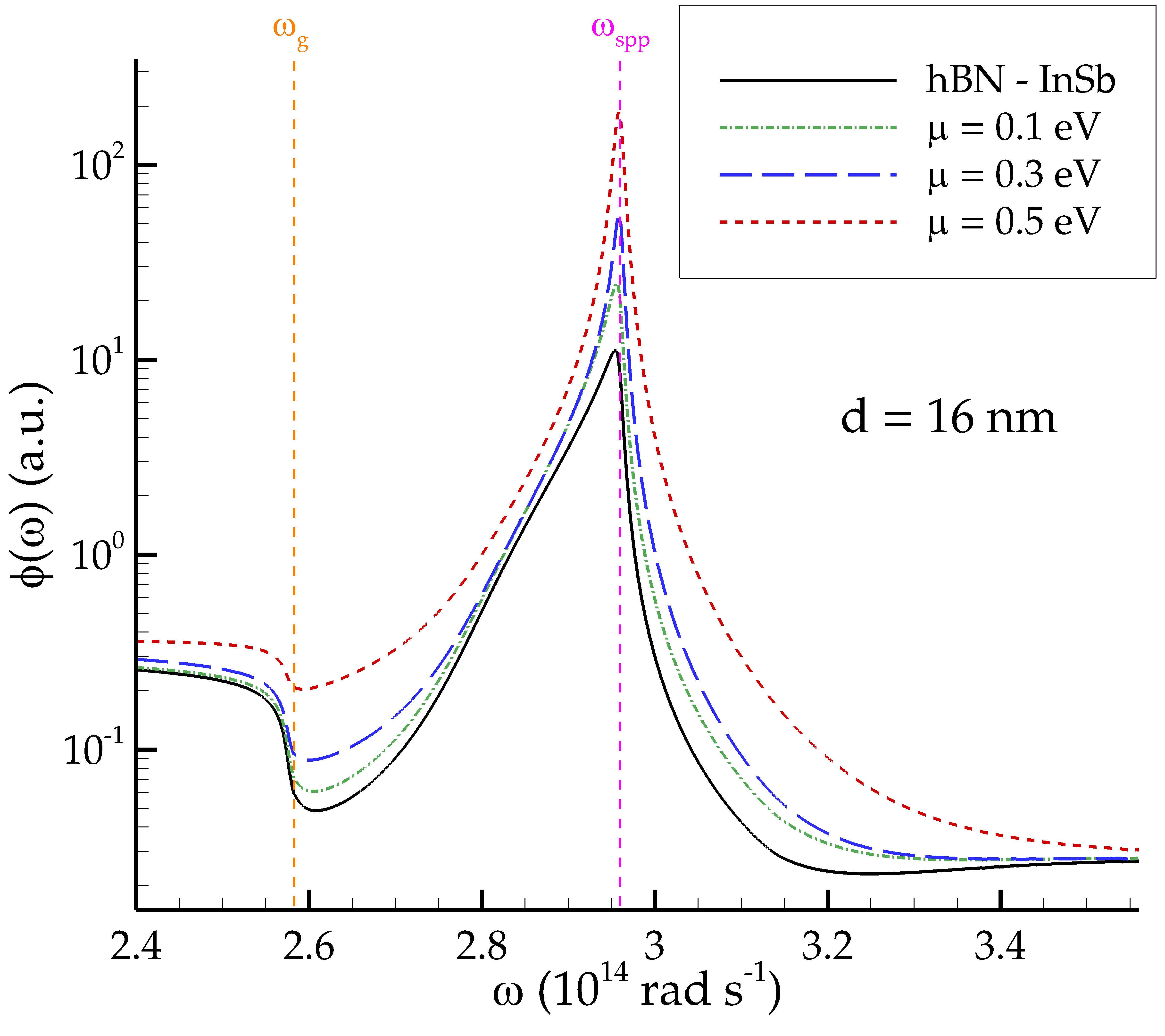}}
\caption{Spectral distribution of radiative flux. The spectral flux $\phi(\omega)$ is represented for $d=16\,$nm in absence of graphene, and with graphene for $\mu=0.1,0.3,0.5\,$eV. The curves show an amplification, for any value of the chemical potential, at $\omega=\omega_\text{spp}$. Moreover, for $\mu=0.5\,$eV the width of the peak is significantly enhanced, due to the coupling of surface resonance modes.}
\label{FigSpectrum}\end{figure}

We finally discuss the effect of graphene on the flux spectrum $\phi(\omega)$ (\ref{FigSpectrum}). The first manifest effect is the amplification, for any value of the chemical potential, of the peak of the spectrum at $\omega=\omega_\text{spp}$. This clearly corresponds to the enhancement of the cutoff wavevector $k_c$ (i.e. an increased number of participating modes) observed for any considered value of $\mu$. Moreover, the coupling discussed for $\mu=0.5\,$eV is here reproduced in terms of enlargement of the peak. These curves clearly explain that both the radiation exchange and the electric power are amplified. Nevertheless, they also allow to explain the enhancement of efficiency. As a matter of fact, the modifications to the spectral properties are more pronounced around $\omega_\text{spp}$, which is larger than $\omega_g$. Thus, since the region $\omega<\omega_g$ contributes only to $P_\text{rad}$, this amplification results in a higher value of $\eta$ as well.

We have proposed a novel setup for NTPV energy conversion in which the PV cell is covered with a graphene sheet. The presence of this two-dimensional system modifies the optical response of the cell, causing in particular the appearance of new surface resonant modes which coupling to the ones belonging to the source produce a significant enhancement of the electric power produced in the cell as well as of the overall efficiency of the NTPV cell. These graphene-based photovoltaic cells have a considerable potential to enable ultrahigh-efficiency electricity production from thermal energy.

\section*{Appendix}

We give here a brief overview of the optical properties of graphene used throughout the calculation presented in the paper. The response of the graphene sheet is described in terms of a 2D frequency-dependent conductivity $\sigma(\omega)$, written as a sum of an intraband (Drude) and an interband contribution respectively given by \cite{FalkovskyJPhysConfSer08}
\begin{equation}\begin{split}\sigma_D(\omega)&=\frac{i}{\omega+\frac{i}{\tau}}\frac{2e^2k_BT}{\pi\hbar^2}\log\Bigl(2\cosh\frac{\mu}{2k_BT}\Bigr),\\
\sigma_I(\omega)&=\frac{e^2}{4\hbar}\Biggl[G\Bigl(\frac{\hbar\omega}{2}\Bigr)+i\frac{4\hbar\omega}{\pi}\int_0^{+\infty}\frac{G(\xi)-G\Bigl(\frac{\hbar\omega}{2}\Bigr)}{(\hbar\omega)^2-4\xi^2}\,d\xi\Biggr]\end{split}\end{equation}
where $G(x)=\sinh(x/k_BT)/[\cosh(\mu/k_BT)+\cosh(x/k_BT)]$. The conductivity also depends on the temperature $T$ of the graphene sheet (assumed equal to the temperature of the cell), on the chemical potential $\mu$ and on the relaxation time $\tau$, for which we have chosen the value $\tau=10^{-13}\,$s \cite{JablanPRB09}.

In order to deduce the heat flux between the source and the graphene-covered cell we need the expression of the reflection coefficients for an arbitrary frequency $\omega$, wavevector $\mathbf{k}$ and polarization $p$. To this aim we assume that the graphene sheet is located in $z=z_0$ and that it separates two non-magnetic media 1 ($z<z_0$) and 2 ($z>z_0$) having dielectric permittivities $\varepsilon_1(\omega)$ and $\varepsilon_2(\omega)$ respectively. We assume the presence of an incoming field for $z<z_0$, generating a reflected field in medium 1 and a transmitted field in medium 2. We then impose the continuity of component of the electric field parallel to the graphene sheet and connect the discontinuity of the magnetic field to the surface current on the sheet, proportional to the electric field on the sheet through the conductivity $\sigma(\omega)$. This procedure gives the following expressions of the reflection and transmission coefficients for a given couple $(\omega,\mathbf{k})$ and for the two 
polarizations
\begin{equation}\begin{split}r_\TE&=e^{2ik_z^{(1)}z_0}\frac{k_z^{(1)}-k_z^{(2)}-\mu_0\sigma(\omega)\omega}{k_z^{(1)}+k_z^{(2)}+\mu_0\sigma(\omega)\omega},\\ t_\TE&=e^{i(k_z^{(1)}-k_z^{(2)})z_0}\frac{2k_z^{(1)}}{k_z^{(1)}+k_z^{(2)}+\mu_0\sigma(\omega)\omega},\\ r_\TM&=e^{2ik_z^{(1)}z_0}\frac{\varepsilon_2(\omega)k_z^{(1)}-\varepsilon_1(\omega)k_z^{(2)}+\frac{\sigma(\omega)k_z^{(1)}k_z^{(2)}}{\varepsilon_0\omega}}{\varepsilon_2(\omega)k_z^{(1)}+\varepsilon_1(\omega)k_z^{(2)}+\frac{\sigma(\omega)k_z^{(1)}k_z^{(2)}}{\varepsilon_0\omega}},\\ t_\TM&=e^{i(k_z^{(1)}-k_z^{(2)})z_0}\frac{2\varepsilon_1(\omega)k_z^{(2)}}{\varepsilon_2(\omega)k_z^{(1)}+\varepsilon_1(\omega)k_z^{(2)}+\frac{\sigma(\omega)k_z^{(1)}k_z^{(2)}}{\varepsilon_0\omega}},\end{split}\end{equation}
where
\begin{equation}k_z^{(i)}=\sqrt{\varepsilon_i(\omega)\frac{\omega^2}{c^2}-\mathbf{k}^2}\end{equation}
is the $z$ component of the wavevector inside medium $i$ ($i=1,2$). For the purpose of our calculation we simply have to take $\varepsilon_1(\omega)=1$, obtaining
\begin{equation}\begin{split}r_\TE&=e^{2ik_zz_0}\frac{k_z-k_z^{(2)}-\mu_0\sigma(\omega)\omega}{k_z+k_z^{(2)}+\mu_0\sigma(\omega)\omega},\\
t_\TE&=e^{i(k_z-k_z^{(2)})z_0}\frac{2k_z}{k_z+k_z^{(2)}+\mu_0\sigma(\omega)\omega},\\
r_\TM&=e^{2ik_zz_0}\frac{\varepsilon_2(\omega)k_z-k_z^{(2)}+\frac{\sigma(\omega)k_zk_z^{(2)}}{\varepsilon_0\omega}}{\varepsilon_2(\omega)k_z+k_z^{(2)}+\frac{\sigma(\omega)k_z^{(1)}k_z^{(2)}}{\varepsilon_0\omega}},\\ t_\TM&=e^{i(k_z-k_z^{(2)})z_0}\frac{2k_z^{(2)}}{\varepsilon_2(\omega)k_z+k_z^{(2)}+\frac{\sigma(\omega)k_zk_z^{(2)}}{\varepsilon_0\omega}}.\end{split}\end{equation}

\begin{acknowledgments}
The authors thank M. Antezza for fruitful discussions. P. B.-A. acknowledges the support of the Agence Nationale de la Recherche through the Source-TPV project ANR 2010 BLANC 0928 01.
\end{acknowledgments}

\end{document}